\documentclass[aip,jcp,a4paper,twocolumn, showkeys]{revtex4}

\usepackage[]{amsmath}
\usepackage{amssymb}
\usepackage[dvips]{graphicx}
\usepackage{color}
\usepackage{tabularx}
\usepackage{mathtools}
\usepackage{algpseudocode}
\usepackage{enumitem}
\usepackage{multirow}
\usepackage{epstopdf}  
\usepackage{bm}  


\newcommand{\vect}[1]{\textbf{\textit{#1}}}

\newcommand{\deep}{\mathrm {D}}
\newcommand{\classical}{\mathrm {C}}
\newcommand{\classicalb}{\mathrm {C,b}}
\newcommand{\classicalnb}{\mathrm {C,nb}}
\newcommand{\trans}{\mathrm {T}}
\newcommand{\thf}{\mathrm {T}}
\newcommand{\methodname}[0]{{AMM}}

\begin{document}

\title{
%Adaptive modeling method for concurrent coupling of two molecular models
%  Adaptive modeling:
Adaptive coupling of a deep neural network potential to a classical force field
}
\author{Linfeng Zhang}
\affiliation{Program in Applied and Computational Mathematics, Princeton University, Princeton, NJ 08544, USA}
\author{Han Wang}
\email{wang\_han@iapcm.ac.cn}
\affiliation{Laboratory of Computational Physics,
  Institute of Applied Physics and Computational Mathematics, Huayuan Road 6, Beijing 100088, P.R.~China}
\author{Weinan E}
\email{weinan@math.princeton.edu}
\affiliation{Department of Mathematics and Program in Applied and Computational Mathematics, Princeton University, Princeton, NJ 08544, USA}
\affiliation{Center for Data Science and Beijing International Center for Mathematical Research, Peking University, P.R.~China}
\affiliation{Beijing Institute of Big Data Research, Beijing, 100871, P.R.~China}

\begin{abstract}
An adaptive modeling method (\methodname{}) that couples a deep neural network potential and a classical force field is introduced to address the accuracy-efficiency dilemma faced by the molecular simulation community.
The \methodname{} simulated system is decomposed into three types of regions. 
The first type captures the important phenomena in the system and requires high accuracy, 
for which we use the Deep Potential Molecular Dynamics (DeePMD) model in this work.
The DeePMD model is trained to accurately reproduce the statistical properties of the \emph{ab initio} molecular dynamics.
The second type does not require high accuracy and a classical force field is used to describe it in an efficient way.
The third type is used for a smooth transition between the first and the second types of regions.
By using a force interpolation scheme and imposing a thermodynamics force in the transition region, 
we make the DeePMD region embedded in the \methodname{} simulated system as if it were embedded in a system that is fully described by the accurate potential. 
A representative example of the liquid water system is used to show the feasibility and promise of this method.
%\bluec{Han: Here I use DeePMD, because the DNN is not discussed in the manuscript.}
\end{abstract}

\maketitle

\section{Introduction}
The molecular simulation community are often faced with the accuracy-efficiency dilemma:
the atomic interaction in the \emph{ab initio} molecular dynamics (AIMD) \cite{marx2000ab} is accurately modeled by the density functional theory (DFT) \cite{hohenberg1964inhomogeneous,kohn1965self,martin2004electronic},
but the extensive computational cost that typically scales cubically with respect to the number of atoms limits its applications to system size of a few hundreds of atoms and simulation time of a few hundreds of picoseconds.
On the other hand, molecular dynamics (MD) with atomic interaction modeled by classical force fields (FFs) can easily scale to millions of atoms, but the accuracy and transferability of classical FFs is often in question.

For a large class of MD applications, people have been addressing this dilemma by multi-scale modeling.
In these applications,  only the accuracy of part of the system is crucial to the phenomena of interest.
Taking the problem of protein folding as an example, the accuracy of modeling the interactions within protein atoms and the interaction between the protein and nearby solvent molecules
dominates the conformation of the protein and the folding process.
Therefore, a natural idea of saving the computational cost while preserving the accuracy is to model the protein and nearby water molecules by an accurate but presumably expensive model, 
whereas to model other water molecules by a cheaper model.
Methods of particular interest are the hybrid quantum mechanics/molecular mechanics (QM/MM) approach \cite{warshel1976theoretical,lin2007qm}, which combines QM models and classical molecular models, 
and the adaptive-resolution-simulation (AdResS) technique \cite{praprotnik2005adaptive,delle2017molecular}, which combines atomic models and coarse-grained models.
{
  It is noted that interpolating the force from different models is commonly used in the multi-scale modeling methods,
  for example, the force-mixing QM/MM method~\cite{bernstein2009hybrid,bernstein2012qm,varnai2013tests,mones2015adaptive},
  the force interpolation AdResS for classical~\cite{praprotnik2005adaptive,fritsch2012adaptive} or path-integral MD~\cite{poma2010classical,agarwal2015path},
  and the force-blending atomistic-continuum coupling methods~\cite{li2012positive,lu2013convergence,li2014theory,wang2018posteriori}.
}

Recently, machine learning (ML) methods have brought in another solution to this dilemma~\cite{behler2007generalized,morawietz2016van,bartok2010gaussian,rupp2012fast,schutt2017quantum,chmiela2017machine,smith2017ani,han2017deep,zhang2018deep}. 
After fitting the AIMD data, these approaches target at an accurate and much less expansive potential energy surface,
thereby eliminating the need of calculating electronic structure information on the fly.
A representative example is the Deep Potential Molecular Dynamics method (in abbreviation, DeePMD)~\cite{han2017deep,zhang2018deep} that the authors recently developed with collaborators.
In this scheme the many-body potential energy is a sum of the ``atomic energies'' associated to the individual atoms in the system.
Each one of these ``atomic'' energies depends analytically, via the deep neural network representation, on the symmetry-preserving coordinates of the atoms belonging to the local environment of each given atom. 
Upon training, DeePMD faithfully reproduces the distribution information of trajectories from AIMD simulations, with nuclei being treated either classically, or quantum mechanically (by path-integral MD).

With the promising features of the ML methods, several problems have motivated us
{to develop an adaptive method that concurrently combines an ML model with a classical FF.
Throughout this paper we use the DeePMD method as an example.
First, accurate training data is expensive and the amount of data is often limited to small number of atoms and conformations.
In addition, for large and complex systems usually we do not have the full QM description but only a part of it. 
Therefore, a practical expectation on the {DeePMD} model should be that it {is trained to be} accurate in the important regions under study, 
whereas the remaining regions could be described by a more simplified classical model.
Second, since the DeePMD model is essentially a many-body potential, the evaluation of energy, force, and virial requires much more floating point operations than classical pairwise potentials.
Taking the liquid water system for example, the DeePMD model is about two orders of magnitudes more expensive than classical TIP3P~\cite{jorgensen1983comparison} water model~\cite{zhang2018deep}.
Therefore, given the same computational resource, the maximal system size that is tractable by the DeePMD model is two orders of magnitudes smaller than that is described by the TIP3P model,
or the longest simulation time achievable is two orders of magnitudes shorter than that of the TIP3P model.
This imposes a limitation on the spacial and temporal scales of the problem if it is only described by the DeePMD model.
Finally, the energy decomposition scheme adopted by the DeePMD model and many other ML methods provides a natural way of doing adaptive modeling. 
This would make the boundary problem in QM/MM due to boundary conditions adopted by electronic computations much less severe.
It should be noted that, electronic structure information, as a natural output of the QM/MM method, will be missing when we perform MD using only an ML model or a classical FF model.
Therefore, we shall limit ourselves to cases that are well described by the potential energy surface 
%(\bluec{Only tow ``PES''s, so I sugguest not using abbreviation}) 
and are less sensitive to electronic degrees of freedoms.

%In some applications of MD simulation, only the accuracy of part of the system is crucial to the phenomena of interest.
%Taking the protein folding for example, the accuracy of modeling
%the interactions within protein atoms and the interaction between the protein and nearby solvent molecules
%dominates the conformation of the protein and the folding process.
%Therefore, a natural idea of saving the computational cost while preserving the accuracy is
%to model the protein and nearby water molecules by DeePMD, whereas model the water molecules far away by a classical model.
%The water molecules should be allowed to free diffuse in the system,
%either closer to or further way from the protein molecule.
%Thus the water model should be allowed to change according to how far the water molecule is away from the protein molecule.
%In other words, the water model should be chosen in an adaptive manner.
In this work, we introduce an adaptive modeling method (\methodname{}) and numerically prove its feasibility in terms of adaptively changing the model for a molecule,  
depending on its spacial position, from the DeePMD model to a classical model, or \emph{vise versa}.
The system is divided into DeePMD regions and classical regions. 
Different regions are bridged by transition regions where the model of a molecule changes smoothly.
The equilibrium between the regions are ensured by the thermodynamic force applying in the transition region.
We demonstrate, by using liquid water as a representative example, that the density profile, the radial distribution functions (RDFs), 
and angular distribution functions (ADFs) in the DeePMD region is in satisfactory agreement with the corresponding subregion of a full DeePMD reference system.
Therefore, the DeePMD region is embedded in the \methodname{} system as if it were embedded in a full DeePMD system.
The statistics of the DeePMD region approximates the grand-canonical ensemble in the thermodynamic limit.

\section{Method}

The \methodname{} simulation region is decomposed into {three types of non-overlapping regions:}
DeePMD regions $\Omega_\deep$ where the many-body atomistic interactions are modeled by the DeePMD scheme \cite{zhang2018deep} ,
classical regions $\Omega_\classical$ where the interactions are modeled by a classical FF model, 
and transition regions $\Omega_\trans$ of uniform thickness $d_\trans$ that bridge the DeePMD regions and the classical regions.
{See an illustrative example in Fig.~\ref{fig:sys}.}
Here we only consider one DeePMD region, one classical region, and the transition region between them.
The case of multiple DeePMD or classical regions can be generalized without substantial difficulty.

We define a reference system, whose only difference with the \methodname{} system is that it is fully described by the DeePMD model in the whole simulation region. 
Our goal is to embed the DeePMD region in the classical region as if it were embedded in a system that is fully modeled by the DeePMD.
In other words, the equilibrium statistical property of the DeePMD region should mimic that of {the corresponding subregion} in the reference system.
{In this sense, since the subregion of the reference system is subject to the grand-canonical ensemble as the size of the system goes to infinity,
the statistics in the DeePMD region approximates the grand-canonical ensemble, and the AMM is a grand-canonical-like molecule dynamics simulation. }

In the \methodname{} scheme, we use a force scheme to fulfill the goal.
We define the force $\vect F_i$ on each atom $i$ as a summation of three components:
\begin{align}\label{eqn:f-total}
  \vect F_i = \vect F_i^{\text{I}} + \vect F_i^{\text{L}} + \vect F_i^{\thf},
\end{align}
where $\vect F_i^{\text{I}}$ is an interpolated force between the DeePMD and the classical model, 
$\vect F_i^{\text{L}}$ is a stochastic force from a Langevin thermostat that controls the canonical distribution,
and $\vect F_i^{\thf}$ is a thermodynamic force that balances the density profile of the whole \methodname{} simulation region. 

In the following we define and discuss in more details the three terms in the definition of $\vect F_i$. 
For the interpolated force $\vect F_i^{\text{I}}$, we define a position dependent characteristic function $w(\bm r)$, which takes a constant value 1 in the DeePMD region and 0 in the classical region,
and changes smoothly from 1 to 0 in the transition region.
The way of defining the characteristic function $w(\bm r)$ 
is not unique, and here we use:
\begin{align}
  w(\bm r) = \left\{
  \begin{aligned}
    & 1 & \bm r &\in \Omega_\deep \\
    & \frac12(1 + \cos\Big[\frac{\pi d(\bm r, \Omega_\deep)}{d_\trans}\Big]) & \bm r & \in \Omega_\trans \\
    & 0 & \bm r &\in \Omega_\classical,
  \end{aligned} \right.
\end{align}
where $d(\bm r, \Omega_\deep) = \min_{\bm s\in\Omega_\deep}\vert\bm r - \bm s\vert$ is the 
{closest distance from the position $\bm r$ to the boundary of the DeePMD region.}
% In the system a weighting function $w(\vect r)$ is used to describe which model (i.e. DeePMD or classical MD) is used at position $\vect r$.
% Therefore, the model describing a molecule may change from DeePMD to classical MD or \emph{vice versa} as the molecule diffuses across the simulation region.
% We adopt the convention that $w = 1$ means the DeePMD is used, while $w=0$ means classical MD is used.
% The weighting function smoothly varies from 1 to 0 between the DeePMD region and the classical MD region.
% We call the region where $0<w<1$ the transition region.
Then $\vect F_i^{\text{I}}$ is defined as a linear interpolation, through $w(\bm r)$, between the DeePMD force and the classical force, i.e.: 
%using the prefactor of the characteristic function $w$, saying
\begin{align}\label{eqn:f-intpl}
  \vect F_i^{\text{I}} = w (\vect R(\vect r_i)) \vect F^\deep_i + [\,1 - w (\vect R(\vect r_i))\,] \vect F^\classical_i
\end{align}
where $\vect F^\deep$ and $\vect F^\classical$ denote the DeePMD force and the classical force, respectively,
and $\vect R(\cdot)$ denotes the characteristic position of atom $i$. 
{In general, $\bm R(\cdot)$ can be directly defined as an identity mapping.
However, in our test example of the water system, we observe that defining $\vect R(\cdot)$ as a mapping from the atomic position to the molecular center-of-mass (COM)
will stabilize the numerical issue caused by the rigidness of the classical water model.
Based on similar considerations, for macromolecules, $\bm R(\cdot)$ can be, e.g., a mapping from atomic positions to the residue COM.}
In addition, we note that instead of a force-interpolation scheme, it is possible to do an energy-interpolation scheme, which conserves the total interpolated energy at the cost of momentum conservation~\cite{delle2007some}. 
The equivalence of the two approaches in terms of equilibrium statistical properties is extensively discussed in Ref.~\cite{wang2015adaptive}.
Here we focus on the force interpolation approach.

The Langevin term $\vect F_i^{\text{L}}$ is defined as
\begin{align}
  \vect F_i^{\text{L}} = -\gamma \vect p_i + \sqrt{m_i} \sigma \dot W,
\end{align}
where $\vect p_i$ and $m_i$ denote the momentum and mass of atom $i$, respectively. 
$W$ denotes the standard Wiener process, the friction $\gamma$ and the noise magnitude $\sigma$ are related by the fluctuation-dissipation theorem $\sigma^2 = 2\gamma k_BT$,
with $k_B$ being the Boltzmann constant and $T$ being the temperature.

The accuracy of the AMM 
is investigated by comparing the statistical properties of the DeePMD region with those of the corresponding region in the full DeePMD reference system.
Due to the difference in the definition of the DeePMD model and the classical model, an imbalance of pressure
%(\bluec{In the context of force interpolation, the density difference is directly caused by pressure difference, while in the context
%of energy interpolation, the density difference is directly caused by chemical potential difference.}) 
on the transition region exists,
which in general will result in an unphysical gap of density profile of different regions, and other higher-order marginal distributions of the configurational distribution functions.
%The accuracy of this grand-canonical-like approximation 
This can be systematically improved by requiring the marginal probability distributions of different orders 
in the $transition$ region identical to those in the full DeePMD model~\cite{wang2013grand}.
The first-order marginal distribution, the density profile, is corrected by the one-body thermodynamic force $\vect F_i^{\thf}$~\cite{fritsch2012adaptive}.
In practice, $\vect F^\thf$ is computed by an iterative scheme:
\begin{align}\label{eqn:cal-thf}
  \vect F^\thf_{k+1} (\vect R) = \vect F^\thf_{k} (\vect R) - \frac{\alpha}{\kappa\rho^2} \nabla \rho_{k}(\vect R),
\end{align}
where $\rho$ denotes the equilibrium number density, $\rho_{k}(\vect R)$ denotes the density profile at the $k$-th iteration step,
$\kappa$ denotes the isothermal compressibility, and $\alpha$ denotes a damping prefactor.
By using the iterative scheme of Eq.~\eqref{eqn:cal-thf}, the converged thermodynamic force will lead to a flat density profile in the system,
which indicates the equilibration between the DeePMD region and the classical MD region.
Higher-order corrections in the transition region, e.g. the correction of the radial distribution function (RDF),
are possible by using the RDF correction to the transition like that proposed in Ref.~\cite{wang2012adaptive}.
In this work, we do not consider the RDF and higher orders corrections, 
and demonstrate, by the numerical example, that the properties in the DeePMD region is satisfactorily accurate by only using the thermodynamic force correction.

\section{Simulation protocol}
\begin{figure}
  \centering
  \includegraphics[width=0.45 \textwidth]{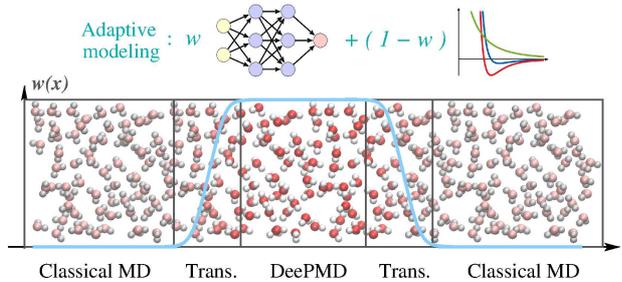}
  \caption{Schematic plot of an adaptive model system.
    From left to right, are the classical, transition, DeePMD, transition and classical regions.
    The blue curve presents the shape of the characteristic function $w(\bm r)$.
  }
  \label{fig:sys}
\end{figure}

In this work the \methodname{} scheme is demonstrated and validated by a water system.
In total 864 water molecules are simulated in a cubic cell of size $7.4668\textrm{nm} \times 1.8667\textrm{nm} \times 1.8667\textrm{nm}$ and subject to periodic boundary conditions.
As shown in Fig.\ref{fig:sys}, the model only changes along $x$ axis.
In a copy of the simulation cell, the DeePMD region of width 2.0~nm locates at the center of the simulation region $x_c = 3.7334$~nm, 
and the thickness of the transition region is $d_\trans=0.3$~nm.
It is noted that this thickness is smaller than those usually used in AdResS methods~\cite{praprotnik2005adaptive,delle2017molecular}, i.e.~twice of the cutoff radius.
Therefore, on average there are roughly 115, 70, and 678 water molecules in the DeePMD region, the transition region, and the classical region, respectively.
The damping prefactor and the compressibility in Eq.~\eqref{eqn:cal-thf}
 are set to 0.25 and $4.6\times 10^{-5}\ \textrm{Bar}^{-1}$, respectively.
% {The thermodynamic force applied in the transition region is determined by the iterative scheme~\eqref{eqn:cal-thf}.}
% \bluec{(Linfeng: I don't think we need to repeat this here. Instead, should we report the parameters in~\eqref{eqn:cal-thf}? The $\alpha$, the $\kappa$, and/or the density.)}
The rest of the system belongs to the classical region, in which the water molecules are modeled by a flexible SPC/E force field~\cite{berendsen1987missing}.
The details of the force field are provided in Appendix~\ref{app:spce}.
In this work,
the atoms are modeled by point-mass particles in both of the DeePMD and classical water models.
% both of the DeePMD and classical water models assumes that the atoms are particles,
% and the generalization  to path-integral MD simulations, in which atoms are model by polymer-rings, is straightforward.
The whole system is coupled to a Langevin thermostat of lag-time 0.1~ps (as a rule of thumb, the friction is set to 10~$\mathrm{ps}^{-1}$~\cite{gao2016sampling}) to keep the temperature at 330~K. 

% {(This paragraph below seems not to belong to a general theory but a practical and technical issues. For example, in some cases the FF could be a rigid model, which is common and we might need a more several compromise. How about moving this para to the next subsection?)}
One practical but important issue is that, the equilibrium covalent bond length of the DeePMD model is not identical to that of the classical force field.
When a molecule leaves the DeePMD region and enters the transition region,
the classical force field switches on, and the mismatched bond length will lead to a bond force with large magnitude.
This may cause the difficulty of equilibrating the bond length in the transition region.
One solution is to use a small enough time step,
so the prefactor $w (\vect R(\vect r_i))$ is small to suppress the large bond force as the molecule enters the transition region.
Another solution, which we use in this work for the simulation efficiency, is to slightly modify the force interpolation scheme~\eqref{eqn:f-intpl} as
\begin{align}\nonumber
  \tilde{\vect F_i}^{\text{I}} &= w (\vect R(\vect r_i)) \vect F^\deep_i 
  + [\,1 - w (\vect R(\vect r_i))\,] \vect F^\classicalnb_i  \\ \label{eqn:f-intpl-p}
  &+ \max\{ \varepsilon_p, 1 - w (\vect R(\vect r_i))\,\} \vect F^\classicalb_i, 
\end{align}
where $\vect F^\classicalnb$ and $\vect F^\classicalb$ are the non-bonded and bonded contributions to the classical force field, respectively.
$\varepsilon_p$ is the shape protection parameter, and we take $\varepsilon_p = 0.01$ in this work, {if not stated otherwise}.
With the shape protection parameter, the force in the DeePMD region is thus modified as $\vect F^\deep_i + \varepsilon_p \vect F^\classicalnb_i$,
so the molecular shape of the classical force field is partially preserved in the DeePMD region,
thus  the equilibration of the bond length is much easier when the molecule enters the transition.
Since the protection parameter $\varepsilon_p$ is small,
the molecular configuration in the DeePMD region is not substantially perturbed.
It is noted that in the numerical example,
we do not observe any difficulty of equilibrating the covalent bonds when a molecule leaves the classical region and enters the transition region,
because the intramolecular part of the DeePMD interaction is much softer than the classical force field.

The data for training the  DeePMD water model was generated by a 330~K NVT AIMD simulation  of a 64-molecule bulk water system
with PBE0+TS exchange-correlation functional under periodic boundary condition.
The total length of the AIMD simulation is 20~ps, and the information of the system was saved every time step of 0.5~fs, 
thus 40,000 snapshots of the system is available.
Among the data, 95\%, i.e.~38,000 snapshots,  are used as the training data, while the rest 2,000 snapshots are used as testing data.
The cut-off radius of the DeePMD model is 0.6~nm.
The descriptors (network input) contain both the angular and radial information of 16 closest oxygen atoms and 32 closest hydrogen atoms.
The descriptors contain only the radial information for the rest of neighbors
in the cut-off radius.
% The only the radial information of the rest neighbors within the cut-off radius is provided in the descriptors.
The deep neural network that model the many-body atomic interaction has 5 hidden layer,
each of which has 240, 120, 60, 30, 10 neurons from the input side to the output side, respectively.
The model is trained using the DeePMD-kit package~\cite{WANG2018}.
The detailed description of the training process is available in Ref.~\cite{WANG2018}.
At the end of the training, the root-mean-square errors of the energy and force evaluated by the testing set are 
% {$2.8\times 10^{-2}$~eV (2.7~kJ/mol) (I found that energy per atom or per molecule is more commonly used)}
0.44~meV (i.e.~0.042~kJ/mol, normalized by the number of molecules)
and $2.4\times 10^{-2}$~eV/\AA~(23~kJ/mol/nm), respectively.

\section{Result and discussion}

Since our goal is to make the DeePMD region in the \methodname{} system as if it were embedded in a full DeePMD system,
the essential check should be made by comparing the configurational probability density of the DeePMD region of the \methodname{} system with that of the
corresponding subregion of a full DeePMD reference system. 
The high-dimensional configurational probability density defined in the phase space can not be easily compared,
but the marginal probability densities can be directly computed from MD trajectories, and compared with those from the reference system.
The agreement of the  first-order marginal probability density is checked by the density profile along the $x$-axis, because the system is homogeneous on the $y$- and $z$-directions.
The agreement in the second-order marginal probability density is checked by comparing the oxygen-oxygen (O-O), oxygen-hydrogen (O-H), and hydrogen-hydrogen (H-H) RDFs. 
In addition, we check the agreement of the third marginal probability density in terms of the ADFs of oxygen atoms defined
up to several cut-off radii. %~\cite{larini2010multiscale,das2012multiscale}.
In this work, both the  \methodname{} system and the full DeePMD reference system are simulated for 2000~ps.
The first 200~ps of the trajectories are discarded,
and the rest of the MD trajectories are considered to be fully equilibrated.
The configurations of the systems are recorded every 0.1~ps,
and the  density profile, RDFs, and ADFs are computed from these configurations. 

\begin{figure}
  \centering
  \includegraphics[width = 0.45\textwidth]{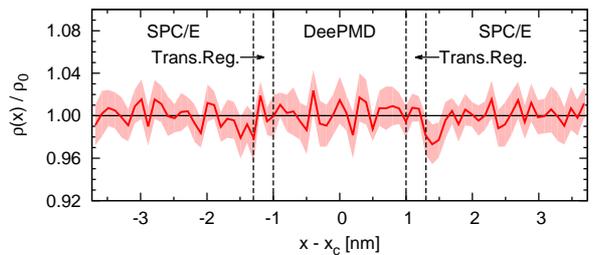}
  \caption{The density profile of the \methodname{} simulation.
    The density is averaged along the $y$ and $z$ directions,
    and the profile is displayed against the $x$ axis.
    The solid line is the density profile,
    and the light shadow denotes the statistical uncertainty of the density profile at the 95\% confidence level.
  }
  \label{fig:dens}
\end{figure}

We report the density profile of the \methodname{} system along the $x$-axis,
and compare it with the equilibrium density of the DeePMD model (denoted by $\rho_0$) in Fig.~\ref{fig:dens}.
The equilibrium profile in the DeePMD region is almost a constant
and  is in satisfactory agreement with the equilibrium density of the full DeePMD result.
The density profiles in the transition regions and in the SPC/E region are also very close to the equilibrium density.
The worst-case deviation, with the statistical uncertainty considered, is 4\% from the equilibrium density.
This indicates that the DeePMD region is embedded in the \methodname{}
system with a similar environment as a full DeePMD system,
in the sense that the density profile of the environment is close to the
equilibrium density of the DeePMD model.

\begin{figure}
  \centering 
  \includegraphics[width = 0.45\textwidth]{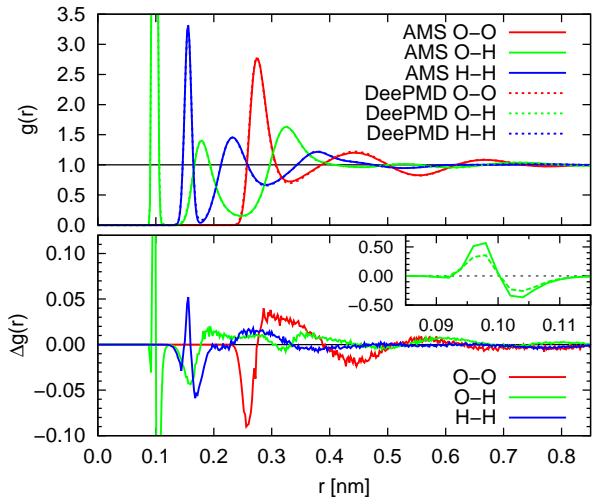}
  \caption{The RDFs of the DeePMD region of the \methodname{} simulation (solid lines) compared with the RDFs of the reference simulation (DeePMD, dotted lines).
    The RDFs of the \methodname{} simulation are presented by solid lines, while those of the reference simulation are presented by dotted lines.
    {The insert of the bottom panel presents the error in the O-H bond length distribution.
    The solid green line shows the case of $d_\trans = 0.3$~nm and $\varepsilon_p = 0.01$,
    while the dashed green line shows the case of $d_\trans = 0.9$~nm and $\varepsilon_p = 0.001$.}
 %   The oxygen-oxygen, oxygen-hydrogen and hydrogen-hydrogen RDFs are plotted by the red, green and blue lines, respectively.
  }
  \label{fig:rdf}
\end{figure}

The O-O, O-H, and H-H RDFs of the DeePMD region in the \methodname{} system is reported in the top panel of Fig.~\ref{fig:rdf}.
All the RDFs are compared with the those computed from the corresponding subregion of the reference system, and
the differences are presented in the bottom panel of Fig.~\ref{fig:rdf}.
The \methodname{} scheme reproduces the intermolecular parts of the RDFs with satisfactory accuracy.
It is noticed that the O-H RDF at around 0.1~nm, which corresponds to the intramolecular O-H bond length distribution, deviates from the reference system.
This deviation is due to the introduction of the shape protection term in the force interpolation~\eqref{eqn:f-intpl-p}. 
In other words, when a water molecule diffuses from the classical region to the DeePMD region, a relaxation time is needed to equilibrate the O-H bond length.
{
  As pointed out by the anonymous referee, it is possible to alleviate the problem by using a larger transition region.
  In this study, we test the case of transition region width $d_\trans = 0.9$~nm,
  which allows a protection parameter that is 10 times smaller, viz.~0.001.
  With this milder shape protection, the O-H bond distribution is restored in a better quality in the DeePMD region
  (see the insertion of the bottom panel of Fig.~\ref{fig:rdf}).
  % It is noted that the protection parameter will eventually vanish as the size of the transition region goes to infinity.
  It is noted that the protection parameter can be further reduced by using a larger transition region,
  however, the computational cost of the \methodname{} simulation will also increase correspondingly.
  This issue will be discussed later in this article.
}

\begin{figure}
  \centering 
  \includegraphics[width = 0.45\textwidth]{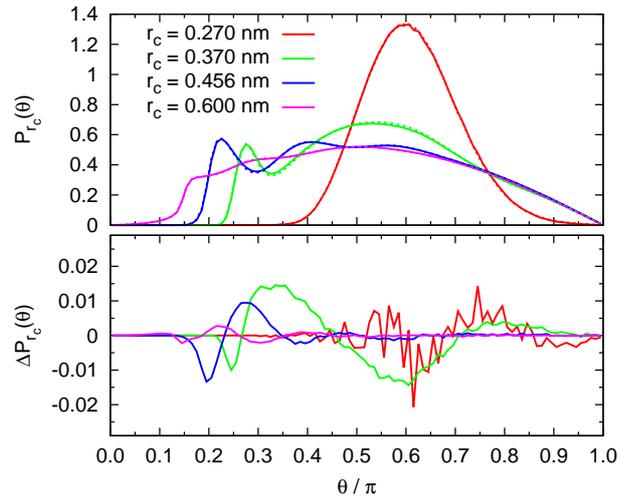}
  \caption{The ADFs of the DeePMD region of the \methodname{} simulation (solid lines) compared with the ADFs of the reference simulation (DeePMD, dotted lines).
    In the top panel, the ADFs of the \methodname{} simulation are presented by solid lines, while those of the reference simulation are presented by dotted lines.
    In the bottom panel, the difference between the ADFs of the \methodname{} and reference simulations are shown.
   % In this figure, difference cutoff radii, 0.270, 0.370, 0.456 and 0.600~nm are plotted by  the red, green and blue lines, respectively.
  }
  \label{fig:adf}
\end{figure}

The ADF is defined by
\begin{align}
  P_{r_c}(\theta) = \frac 1Z
  \Big\langle
  \sum_i \sum_{\substack{ j,k \in \mathcal N(i,r_c)\\  j\neq k}}
  \delta (\theta - \theta_{jik})
  \Big\rangle
\end{align}
where $\mathcal N(i, r_c)$ denotes all the neighboring atoms of $i$ within a cut-off radius $r_c$, 
$\theta_{jik}$ denotes the angle formed by the atoms $j$, $i$ and $k$,
$\langle\cdot\rangle$ denotes the ensemble average,
and $Z$ denotes the normalization factor so that $\int P_{r_c}(\theta) d\theta = 1$.
In Fig.~\ref{fig:adf} we report the oxygen ADF of the DeePMD region of the \methodname{} system
at various cut-off radius $r_c = 0.270$, 0.370, 0.456, and 0.600~nm.
All the results are compared with the reference system, and the differences between the \methodname{}
and the reference system  are presented in the bottom panel of the figure.
The ADFs of the DeePMD region in the \methodname{} system is in satisfactory agreement
with the reference system. 

\begin{figure}
  \centering
  \includegraphics[width = 0.45\textwidth]{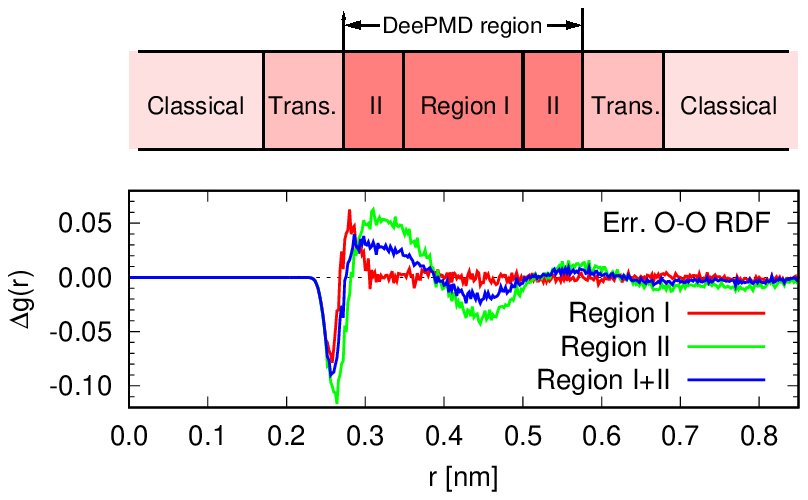}
  \includegraphics[width = 0.45\textwidth]{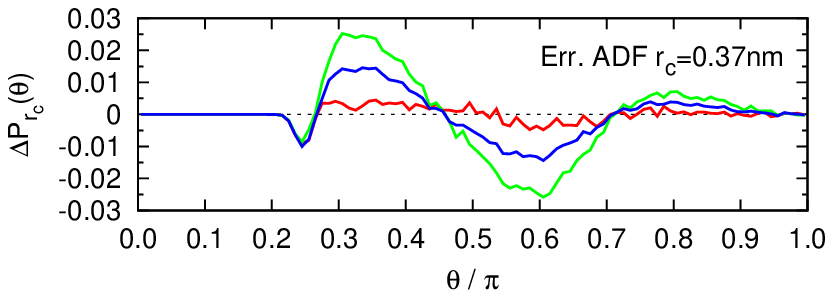}
  \caption{
    {
      The error of the O-O RDF (top panel) and ADF ($r_c$ = 0.37~nm, bottom panel)
      investigated in different subregions of the DeePMD region, i.e.~regions I and II.
      The whole DeePMD region is of width 2.0~nm. The region I is of width 1.0~nm, and the region II is of width 0.5~nm.
    }
  }
  \label{fig:regions}
\end{figure}

{
In addition, we further investigate the RDF and ADF as a function of positions, and 
 plot the error of the RDF and the ADF investigated in different subregions of the DeePMD region.
As shown by Fig.~\ref{fig:regions}, the overall deviations of either the RDF or the ADF compared with the benchmarks are very small.
However, also as expected, the deviation in the region closer to the transition region (Region II) is larger than the deviation lying inside the DeePMD region (Region I).

  The computational cost of AMM simulation is dominated by the computation of the atomic interactions,
  and is estimated by
  $  T = T_\deep + T_\classical$, where $T_\deep$ and $T_\classical$ denotes the computational costs of the DeePMD and classical forces, respectively.
  Since both the DeePMD and classical forces are linearly scalable,
  the costs are estimated by
  $T_\deep = C_\deep(\rho) N_\deep$ and $T_\classical = C_\classical(\rho) N_\classical$,
  where $N_\deep$ and $N_\classical$ denotes the number of atoms on which the DeePMD and classical forces are computed, respectively.
  $C_\deep(\rho)$ and $C_\classical(\rho)$ are density dependent parameters, which are independent with the number of molecules.
  We assume the system is well equilibrated, so that the number of atoms are estimated by
  $N_\deep = \rho (V_\deep + V_\trans + V_B)$, where $V_\deep$ and $V_\trans$ denote
  the volumes of DeePMD and transition regions, respectively.
  The DeePMD force depends on the network derivatives of the neighbors in the cut-off radius $r_c$~\cite{WANG2018},
  therefore, the network derivatives are evaluated for the atoms in a buffering region of width $r_c$ outside the transition region,
  and the computational cost of this part is estimated by $\rho V_B$.
  Similarly, the computational cost of the classical force evaluation is estimated by
  $N_\classical = \rho (V_\classical + V_\trans)$.
  In the end, we have
  \begin{align}\label{eqn:t-esti}
    T_{\mathrm{\methodname{}}} = \tilde C_\deep(\rho) (V_\deep + V_\trans + V_B) + \tilde C_\classical(\rho) (V_\classical + V_\trans)
  \end{align}
  where $\tilde C_\deep (\rho) = \rho C_\deep (\rho)$ and $\tilde C_\classical (\rho) = \rho C_\classical (\rho)$.
  The constants $\tilde C_\deep(\rho)$ and $\tilde C_\classical(\rho)$
  can be estimated by short simulations of small DeePMD and classical systems of the same density.
  It is noted that using a larger transition region improves the accuracy of \methodname{}, but at the same time,
  the extra computational cost grows linearly with respect to the size of the transition region.
  Thus, the size of the transition region should be kept as small as possible,
  as long as the accuracy of \methodname{} is still satisfactory.

  The ratio of the computational cost of the \methodname{} over the full DeePMD simulations is 
  \begin{align}\label{eqn:t-ratio}
    \frac{T_{\mathrm{\methodname{}}}}{T_{\textrm{DPMD}}}
    \approx
    \frac{V_\deep + V_\trans + V_B}{V_{\textrm{sys}}}
    +
    \frac{\tilde C_\classical(\rho)}{\tilde C_\deep(\rho)}
    \frac{V_\trans + V_\classical }{V_{\textrm{sys}}},
  \end{align}
  where $V_{\textrm{sys}} = V_\deep + V_\trans + V_\classical$ is the volume of the whole system.
  % In Eq.~\eqref{eqn:t-ratio}, the approximation
  At the limit of infinitely large classical region, the ratio converges to ${\tilde C_\classical(\rho)}/{\tilde C_\deep(\rho)}$,
  while at the limit of infinitely fast classical force field evaluation,
  the ratio converges to ${(V_\deep + V_\trans + V_B)}/{V_{\textrm{sys}}}$.
  Therefore, ${\tilde C_\deep(\rho)}/{\tilde C_\classical(\rho)}$ and ${V_{\textrm{sys}}}/{(V_\deep + V_\trans + V_B)}$
  are the highest acceleration ration that one obtains from using the \methodname{} method, at the corresponding limits.

  In our example, the constants  $\tilde C_\deep(\rho)$ and $\tilde C_\classical(\rho)$
  are $7.6\times 10^{-3}$~s/nm$^3$/step and $1.2\times 10^{-3}$~s/nm$^3$/step on one core of an Intel Xeon Gold 6148 CPU.
  It is noted that the performance of our in-house code of the classical force field is not optimized.
  As a comparison, the constant of Gromacs 5.1.4~\cite{pronk2013gromacs,abraham2015gromacs} is $7.0\times 10^{-5}$~s/nm$^3$/step,
  which is 17 times faster than the in-house code. 
  The estimated computational time by Eq.~\eqref{eqn:t-esti} is 0.124~s/step,
  while measured wall-time on the same CPU is 0.134~s/step, which validates the estimated~\eqref{eqn:t-esti}.
  The computational cost of a full DeePMD simulation is 0.200~s/step,
  so the \methodname{} saves 33\% (or 38\% by estimate~\eqref{eqn:t-esti}) computational cost of the full DeePMD simulation.
  This may be not significant at the first sight,
  because the volume $V_\deep + V_\trans + V_B$ takes 51\% of the whole system,
  then the \methodname{} cannot save more than 49\% of the computational cost.
  Moreover, the sub-optimized force field code also makes the \methodname{} slower.
  The acceleration of the \methodname{} will be further improved if
  the DeePMD region becomes smaller, the classical region becomes larger,
  or the code of the classical force field is further optimized.
}

\section{Conclusion and perspectives}
In summary, we introduce a promising tool for concurrent coupling of
the DeePMD model and a classical force field.
It should be clear that the same strategy should also be applicable to general cases,
where an expensive model is concurrently simulated with a cheap model.
The requirement for the expensive model is that the part dominating the computational expense is computed in a short-range manner.

% two molecular models.
% In particular, it is useful, for example, to study biomolecules solvated in water, where one could use DeePMD to only parametrize the potential for the biomolecules and nearby water molecules, and couple it to a less expensive water model.
% Even though we only presented an example that combines 
% In this sense, it is possible to treat the long-range part by learning the partial charge with ML method~\cite{artrith2011high},
% and compute the electrostatic interaction with efficient methods like particle mesh Ewald method~\cite{essmann1995spm}.

Future work of this adaptive modeling method is to study biomolecules solvated in water,
where one could use DeePMD to only parameterize the potential for the biomolecules and nearby water molecules, and couple it to a less expensive water model.
It is worth investigating the accuracy in the systems where long-range electrostatic effect plays an important role.
In this situation,
the long-range electrostatic of the ML model should be included by, for example, learning the partial charge based on the atomic environment~\cite{artrith2011high}, and then the point-charge electrostatic is efficiently computed by fast Ewald algorithms~\cite{hockney1988computer,darden1993pme}.
It is also of particular interest to investigate the accuracy of the dynamical properties, like auto-correlation functions,
upon concurrently coupling of different models~\cite{agarwal2015molecular,delle2016formulation}.

\begin{acknowledgments}
The work of LZ and WE is supported in part by ONR grant N00014-13-1-0338, DOE grants DE-SC0008626 and DE-SC0009248, 
and NSFC grants U1430237 and 91530322.
The work of HW is supported by the National Science Foundation 
of China under Grants 11501039, 11871110 and 91530322, 
the National Key Research and Development Program of China 
under Grants 2016YFB0201200 and 2016YFB0201203, 
and the Science Challenge Project No. JCKY2016212A502.
\end{acknowledgments}

\appendix
\section{The flexible SPC/E force field}
\label{app:spce}

\begin{table}
  \centering
  \caption{The parameters used in the flexible SPC/E water model.}
  \label{tab:spce}
  \begin{tabular*}{0.4\textwidth}{@{\extracolsep{\fill}}crr}
    \hline\hline
    Parameter & Value & unit \\ \hline
    $C_{12}$   & $2.6331\times 10^{-6}$ & $\textrm{kJ mol}^{-1} \textrm{nm}^{12}$ \\
    $C_{6}$    & $2.6171\times 10^{-3}$ & $\textrm{kJ mol}^{-1} \textrm{nm}^{6}$ \\
    $q_{\textrm{H}}$ & $0.4238$ & e \\
    $q_{\textrm{O}}$ & $-0.8476$ & e \\
    $r^0_{\textrm{OH}}$ & $0.1$ & nm \\
    $k_{\textrm{OH}}$ & $3.45\times 10^5$ & $\textrm{kJ mol}^{-1}\textrm{nm}^{-2}$ \\
    $\theta^0_{\textrm{HOH}}$ & 109.47 & deg \\
    $k_{\textrm{HOH}}$ & $3.45\times 10^5$ & $\textrm{kJ mol}^{-1}\textrm{rad}^{-2}$ \\
    \hline\hline
  \end{tabular*}
\end{table}

A flexible SPC/E water molecule is modeled by three point-mass particles~\cite{berendsen1987missing}.
The interaction is composed by the non-bonded and bonded parts:
\begin{align}
  U = U_{\textrm{nb}} + U_{\textrm{b}}
\end{align}
The non-bonded interaction has the Coulomb and van der Waals contributions:
\begin{align}
  U_{\textrm{nb}} = U_{\textrm{Coulomb}} + U_{\textrm{vdw}}
\end{align}
Both of the Coulomb and van der Waals contributions are pairwise additive, i.e.~they are of the form $U = \sum_{i\neq j} U(r_{ij})$,
where $r_{ij}$ is the distance between atoms $i$ and $j$.
The Coulomb interaction between an oxygen atom and a hydrogen atom of distance $r$ reads
\begin{align}
  U_{\textrm{Coulomb,OH}}(r) = \frac{q_{\textrm{O}} q_{\textrm{H}}}{4\pi\epsilon_0\epsilon_r} \frac1r
\end{align}
where $q_{\textrm{O}}$ and $q_{\textrm{H}}$ are partial charges of the oxygen and hydrogen atoms, respectively, which are defined in Tab.~\ref{tab:spce}.
The Coulomb interaction of oxygen-oxygen and hydrogen-hydrogen atom pairs are defined analogously.
The oxygen atoms in the system interact with the Lennard-Jones 6-12 potential:
\begin{align}
  U_{\textrm{LJ,OO}} (r) = \frac{C_{12}}{r^{12}} - \frac{C_6}{r^6}
\end{align}
where $r$ is the distance between the oxygens, and $C_{12}$ and $C_6$ are force field parameters given in Tab.~\ref{tab:spce}.
% The electronic structure of the molecule is approximated by the partial point charges  locating on the atoms.
% The magnitude of the partial charge of oxygen and hydrogen are denoted by $q_{\textrm{O}}$ and $q_{\textrm{H}}$, respectively.
The bonded interaction has the bond stretching and the angle bending contributions:
\begin{align}
  U_{\textrm{nb}} = U_{\textrm{bond}} + U_{\textrm{angle}}.
\end{align}
The bond stretching of the O-H covalence bond is modeled by a harmonic potential:
\begin{align}
  U_{\textrm{bond,OH}} (r) = \frac 12  k_{\textrm{OH}} ( r_{\textrm{OH}} - r^0_{\textrm{OH}}) ^2,
\end{align}
where $r_{\textrm{OH}}$ and $r^0_{\textrm{OH}}$ are the the O-H bond length and the equilibrium bond length, respectively, and $k_{\textrm{OH}}$ is the spring constant.
The bending of the H-O-H angle is modeled by a harmonic potential of the angle
\begin{align}
  U_{\textrm{angle,HOH}} (\theta) = \frac 12  k_{\textrm{HOH}} ( \theta_{\textrm{HOH}} - \theta^0_{\textrm{HOH}}) ^2,  
\end{align}
where $\theta_{\textrm{HOH}}$ and $\theta^0_{\textrm{HOH}}$ are the the H-O-H angle and the equilibrium angle, respectively, and $k_{\textrm{HOH}}$ is the spring constant.
The values of the parameters of the flexible SPC/E model is provided in Tab.~\ref{tab:spce}.

% \bibliography{ref}{}

\begin{thebibliography}{10}
\bibitem{lin2007qm}
Hai Lin and Donald~G Truhlar.
\newblock Qm/mm: what have we learned, where are we, and where do we go from
  here?
\newblock {\em Theoretical Chemistry Accounts}, 117(2):185, 2007.
\end{thebibliography}
% \bibliographystyle{unsrt}
% \bibliographystyle{aipauth4-1}

\end{document}